\documentclass[11pt]{article}
\usepackage{amsthm,amsfonts,amsmath,amssymb,amscd, accents, mathrsfs, mathtools}
\usepackage{authblk}

\usepackage[numbers]{natbib}
\usepackage[section]{placeins}
\usepackage{float}
\usepackage{program}
\catcode`\|=12\relax
\usepackage{subfigure}
\usepackage[labelsep=space]{caption}
\usepackage{graphicx}
\usepackage{dcolumn}
\usepackage{bm}
\usepackage{etoolbox} 
\patchcmd{\subequations}{{0}}{{-1}}{}{}
\patchcmd{\subequations}{\alph}{.\arabic}{}{}
\usepackage{amsmath,amsfonts,amssymb,amsthm}
\usepackage{booktabs}

\allowdisplaybreaks

\theoremstyle{thmstyleone}%
\theoremstyle{thmstyletwo}%

\theoremstyle{thmstylethree}%

\author{
Allan Wing-Bocanegra\\
Tecnologico de Monterrey, Escuela de Ingenieria y Ciencias\\
Ave. Eugenio Garza Sada 2501, Monterrey, 64849, N.L., Mexico.\\
  \texttt{A00832476@tec.mx}
  \and
Salvador E. Venegas-Andraca\\
Tecnologico de Monterrey, Escuela de Ingenieria y Ciencias\\
Ave. Eugenio Garza Sada 2501, Monterrey, 64849, N.L., Mexico.\\
  \texttt{svenegas@tec.mx}
}

\title{Unitary Coined Discrete-Time Quantum Walks on Directed Multigraphs}

\begin{document}

\maketitle







\begin{abstract}
Unitary Coined Discrete-Time Quantum Walks (UC-DTQW) constitute a universal model of quantum computation, meaning that any computation done by a general purpose quantum computer can either be done using the UC-DTQW framework. In the last decade,s great progress has been done in this field by developing quantum walk-based algorithms that can outperform classical ones. However, current quantum computers work based on the quantum circuit model of computation, and the general mapping from one model to the other is still an open problem. In this work we provide a matrix analysis of the unitary evolution operator of UC-DTQW, which is composed at the time of two unitary operators: the shift and coin operators. We conceive the shift operator of the system as the unitary matrix form of the adjacency matrix associated to the graph on which the UC-DTQW takes place, and provide a set of equations to transform the latter into the former and vice-versa. However, this mapping modifies the structure of the original graph into a directed multigraph, by splitting single edges or arcs of the original graph into multiple arcs. Thus, the fact that any unitary operator has a quantum circuit representation means that any adjacency matrix that complies with the transformation equations will be automatically associated to a quantum circuit, and any quantum circuit acting on a bipartite system will be always associated to a multigraph. Finally, we extend the definition of the coin operator to a superposition of coins in such a way that each coin acts on different vertices of the multigraph on which the quantum walk takes place, and provide a description of how this can be implemented in circuit form.
\end{abstract}

\section{\label{sec:level1}Introduction}





Quantum walks, originally designed to model quantum phenomena \cite{godoy92,feynman86,gudder88,aharonov93}, are an advanced tool for building quantum algorithms (e.g. \cite{shenvi02,childs2003,ambainis07,kendon06,morley19,latif1901,latif1902,konno08,portugal13,campos2021}) and analyzing biological data \cite{sadiego2020} that has been shown to constitute a universal model of quantum computation \cite{childs09,lovett10}. Quantum walks come in two variants: discrete and continuous in time  \cite{Venegas-Andraca2012}.





A Discrete-Time Quantum Walk (DTQW) consists of a bipartite quantum state, $|\psi \rangle = |v\rangle\otimes |c\rangle$, containing information about the position and coin states of a walker on a graph $\mathcal{G}$, and a unitary \textit{evolution operator}, $U$, composed of the subsequent application of a \textit{coin operator}, $C$, and a \textit{shift operator}, $S$. $U$ has an action on $|\psi \rangle$ such that each time it is applied, it produces transitions between pairs of quantum states associated to adjacent vertices on the graph $\mathcal{G}$ where the DTQW takes place. 


Different approaches have been made to address the problem of constructing evolution operators that are able to generate discrete-time quantum walks on different topologies, e.g.: Aharonov $et\; al.$ \cite{aharonov_2002} proposed the Reversible Arc model, Montanaro \cite{montanaro_2007} proposed the Directed-Graph model, Portugal $et\; al.$ \cite{portugal_staggered} proposed the Staggered Quantum Walk model, Szegedy \cite{szegedy_2004} proposed a model known by the author's name, Attal $et\; al.$ proposed the Open Quantum Random Walk model \cite{open_quantum_walks}, and Zhan \cite{godsil_zhan_2019} proposed a model called Vertex-Face Walk. Nevertheless, there are two problems with the evolution operators proposed in the existing models we will study in this work:

\begin{itemize}
    \item 
    Generally speaking, the design of evolution operators for running quantum walks on arbitrary topologies is a difficult task.

    \item
    Current quantum computer technology is based on the circuit model of quantum computation. Although it is known that the models of gate-based quantum computation and quantum walks are both universal, it remains unknown how to efficiently transform a quantum-walk based algorithm into a quantum gate-based algorithm and vice versa. 
       
\end{itemize}


In this work, we consider the shift operator of a unitary quantum walk $S$ as a unitary matrix with the information of the connections between the vertices of the graph on which we want to perform a quantum walk, similar to a left stochastic transition matrix in a random walk \cite{portugal13}. To induce the connectivity information in $S$, we provide a system of equations with which we can map the transpose adjacency matrix, $\mathcal{A}^{\intercal}$, of a certain graph, $\mathcal{G}$, into $S$ -- $\mathcal{A}^{\intercal}$ is an unnormalized version of the left stochastic transition matrix in a random walk. This tackles the first problem by standardizing the construction of the evolution operator of a quantum walk. However,  even though $S$ and $\mathcal{A}^{\intercal}$ contain the same connectivity information, $S$ is not associated to exactly the same graph $\mathcal{G}$ as $\mathcal{A}^{\intercal}$. Instead, $S$ is associated to $\mathcal{G}'$, which is a mapping of $\mathcal{G}$ into a multigraph that complies with unitarity conditions, that we will describe in the next chapter. This model can be seen as an extension of the Directed-Graph model proposed by Montanaro \cite{montanaro_2007}.

The proposed mapping comes from the fact that the quantum system we use to perform a DTQW is bipartite, i.e. $|\psi\rangle = |c\rangle\otimes|v\rangle$, and any operator acting on a bipartite system has a block matrix form. In \cite{petz_1970}, Petz proves this property for unitary operators and shows that the block matrices that compose them are indeed Kraus operators, and that each column of the unitary block matrix forms a different set of Kraus operators. In this work, we will use this property to provide a link between the shift operator of a unitary coined discrete-time quantum walk, usually referred to as DTQW for simplicity, and the adjacency matrix of the graph, $\mathcal{G}$, on which the quantum walk takes place, as well as a detailed description of the dynamics of a quantum walker on $\mathcal{G}$.

The fact that any operator acting on a bipartite system has a block matrix form, has also been used to develop the theory of Open Quantum Walks on \cite{open_quantum_walks,sinayskiy_petruccione_2019,  kemp_sinayskiy_petruccione_2020}, which, as the name implies, use an Open Quantum System to represent the quantum state of a walker, although, in this case the evolution operator is not necessarily unitary. Unitarity conditions can be applied to the evolution operator of an open quantum walk to obtain a unitary quantum walk, although, given that the open quantum walk model does not contain a coin operator, we will obtain a unitary coinless discrete-time quantum walk. That is to say, the evolution operator of an Open Quantum Walk is an analog to the shift operator and not to the complete evolution operator described in this work. 

To address the second problem associated to current evolution operators, we consider the inverse mapping used to create a multigraph $\mathcal{G}'$ out of a general graph $\mathcal{G}$, and the inverse mapping used to create $S$ out of $\mathcal{A}^{\intercal}$. This leads to the fact that any unitary operator on a bipartite system can be used as the shift operator of a DTQW. And given that any quantum circuit has a unitary matrix form, we conclude that any quantum circuit acting on a bipartite quantum register can be suitable for perform a DTQW on a general purpose quantum computer. The converse problem, i.e. constructing a quantum circuit given an evolution operator is not addressed in this work.

Finally, regarding the coin operator, we extend its form to a superposition of $n$ coin operators, each acting on one of the $n$ vertices of the multigraph on which the DTQW takes place. We do not restrict the form of the coin operators, i.e. they can be any unitary matrix acting on the coin register. This extension might be useful in cases in which we want the evolution operator to have a different behavior on selected vertices.

\section{Basic Definitions on Graph Theory}

In this section we present basic definitions about the type of graphs we use in this paper. The following definitions are mainly taken from \cite{xu_2003}.

\begin{figure}[htbp]
\centering

\begin{minipage}[t]{.4\textwidth}
\subfigure[]{
\resizebox{\textwidth}{!}{    
\includegraphics{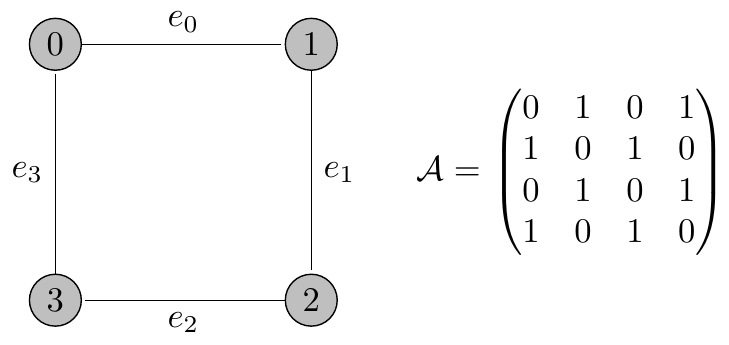}
}
\label{graph_a}
}
\end{minipage}
\begin{minipage}[t]{.4\textwidth}
\subfigure[]{
\resizebox{\textwidth}{!}{    
\includegraphics{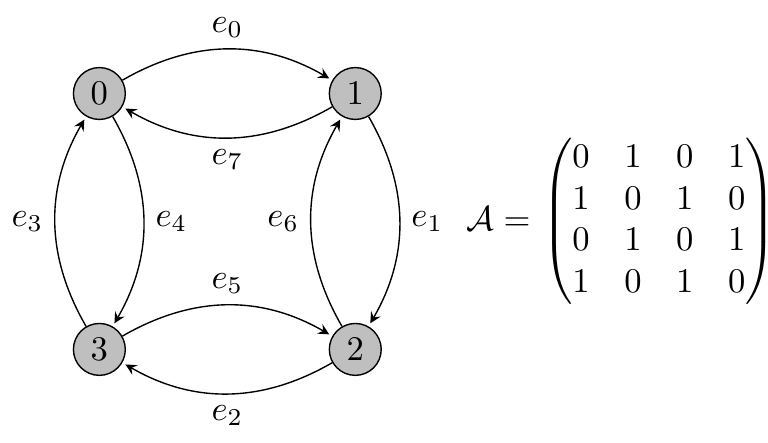}
}
\label{graph_b}
}
\end{minipage}

\begin{minipage}[t]{.4\textwidth}
\subfigure[]{
\resizebox{\textwidth}{!}{    
\includegraphics{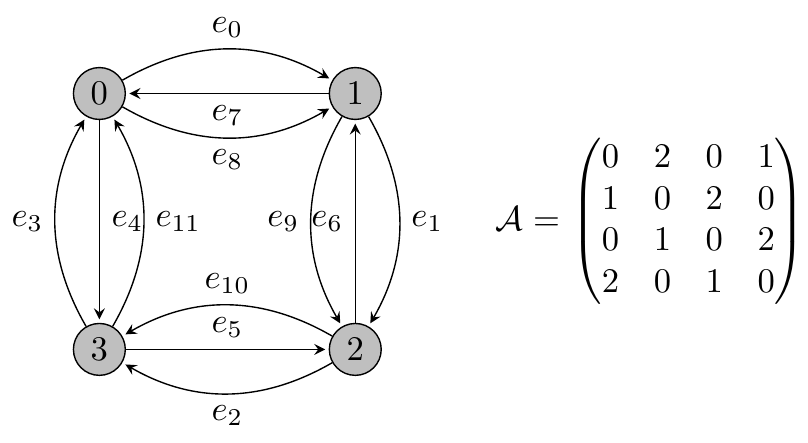}
}
\label{graph_c}
}
\end{minipage}
\begin{minipage}[t]{.4\textwidth}
\subfigure[]{
\resizebox{\textwidth}{!}{    
\includegraphics{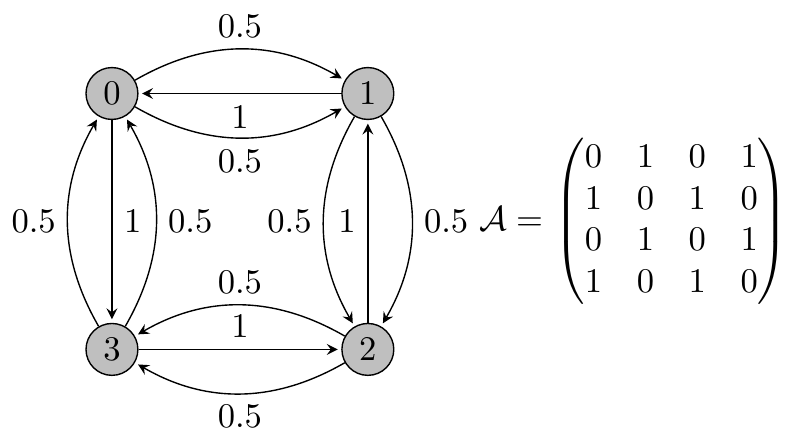}
}
\label{graph_d}
}
\end{minipage}

\caption{Examples of different types of graphs along with their adjacency matrix representation. (a) Undirected graph. (b) Directed graph. (c) Multigraph. (d) Weighted graph. Graphs (a), (b) and (c) are unweighted and labeled with the element $e_k$ that distinguishes each edge, while graph (d) is weighted and labeled with the weight $w(e_k)$ associated to each edge. Notice that the adjacency matrix is the same for graphs (a), (b) and (d), although their connections differ. This means the same adjacency matrix can represent multiple graphs.}
\label{types_of_graphs}
\end{figure}

{\bf Definition 1.} A graph is an ordered triple $(V,E,f)$, where $V$ is a non-empty set called the $vertex\; set$, $E$ is a set called $edge\; set$, and $f:E\rightarrow V \times V$ is a mapping, called $incidence\; function$, which maps an edge $e_k\in E$ into an ordered or unordered pair of vertices, i.e. $(v_i,v_j)$ or $\{v_i,v_j\} \in V\times V$, respectively, called $end\; vertices$. The sets $E$ and $V$ are disjoint. This definition allows loops, i.e. the end vertices can be the same vertex for a given edge.

{\bf Definition 2.} An undirected graph is a graph for which $V\times V$ is a set of unordered pairs $\{v_i,v_j\}$. These graphs are called $undirected\; edges$.

{\bf Definition 3.} A $directed\; graph$ or $digraph$ is a graph for which $V\times V$ is a set of ordered pairs $(v_i,v_j)$, i.e. $f(e_k)=(v_i,v_j)$. Then $e_k$ is called a $directed\; edge$ or $arc$ and, $v_i$ and $v_j$ are called, $tail$ and $head$ of the arc $e_k$, respectively.

{\bf Definition 4.} A multigraph is a graph for which more than one edge can be mapped into the same element $V\times V$ under $f$.

{\bf Definition 5.} A directed multigraph is a digraph that allows parallel arcs.

{\bf Definition 6.} A $weighted\; graph$ is a graph whose edges are assigned and labeled with a numerical value, denoted by $w(e_k)$. An $unweighted\; graph$ is a graph for which $w(e_k)=1\; \forall \; e_k\in E$.

{\bf Definition 7.} The $indegree$ of a vertex $v_i\in V$ is the number of arcs coming into $v_i$. Similarly, the $outdegree$ of a vertex $v_i$ is the number of arcs going out of $v_i$. The $degree$ of $v_i$ is the sum of in- and out- degrees of $v_i$.


{\bf Definition 8.} 
An adjacency matrix is the matrix representation of a graph, and is denoted by $\mathcal{A}$, where $a_{ij} \in \mathcal{A}$ is the sum of the weights of all directed edges with tail $v_i$ and head $v_j$. If undirected edges connect $v_i$ and $v_j$, their weights will be added to both $a_{ij}$ and $a_{ji}$. This definition enables us to assign multiple graphs to the same adjacency matrix, as can be seen from figures \ref{graph_a}, \ref{graph_b} and \ref{graph_d}.

{\bf Definition 9.} 
A graph is said to be connected if there exists a path of adjacent edges, i.e. edges that have a vertex in common, connecting all pairs of vertices.

\section{Walks on Graphs}
 \label{walks_on_graphs}
\subsection{Classical Random Walks}
Let $G = (V,E)$ be a connected graph with $n$ nodes, arbitrarily labeled from 0 to $n-1$, and $m$ edges for each node. A random walk on $G$ is defined as a process in which a walker starts at a vertex $v_0$, and randomly moves on to one of the adjacent vertices. After $t$ steps, the walker stops at node $v_t$, and the path traced along the graph, i.e. the sequence of random nodes, is a Markov chain. To study the process in vector form, we define the walker's probability vector after $t$ steps as

\begin{equation}
    \label{probability_vector}
    P_t = p^t_0\hat{e}_0 + p^t_1\hat{e}_1 + \dots + p^t_{n-1}\hat{e}_{n-1}=
    \begin{pmatrix}
    p^t_0\\
    p^t_1\\
    \vdots\\
    p^t_{n-1}
    \end{pmatrix}
\end{equation}
Where $p^t_i$ is the probability of finding the walker at vertex $i$ at step $t$ and $\hat{e}_i$ is the standard basis vector of $\mathbb{R}^n$ that contains a 1 in the $i$th entry.

Now let $\mathcal{A}$ be the adjacency matrix of $G$ and $D$ a diagonal matrix that contains the outdegrees of each vertex, i.e.:
\begin{equation}
    D_{ii}=\sum_{j=0}^{n-1} \mathcal{A}_{ij}
\end{equation}
We define the matrix of transition probabilities of this markov chain as 
\begin{equation}
    M = D^{-1}\mathcal{A}
\end{equation}
Thus a random walk can be defined by a simple update rule:
\begin{equation}
    \label{recurrence_random_walk}
    P_{t+1} = M^{\intercal} P_{t}
\end{equation}
and the probability distribution vector after $t$ steps can be calculated in terms of only the transition matrix and the initial probability distribution vector of the system:

\begin{equation}
    \label{t_steps_random_walk}
    P_t = (M^{\intercal})^t P_0
\end{equation}

The reason why we use $M^{\intercal}$ rather than $M$ to generate a random walk is due to directed edges and the relation between the basis vectors $e_i$ and the nodes probabilities $p^t_i$. If we apply $M$ to $P_t$ the probability transitions will correspond to a walk in the inverse direction of the directed edges. $M^{\intercal}$ provides a walk in the correct direction. Different walks might result by associating the probability of node $j$, $p_j$, to basis a vector with a different index, i.e. $e_i$, although in this work we will follow the definition presented in Eq.  \eqref{probability_vector} for simplicity. This is important to mention given that the same rationale will be applied when defining a quantum walk.

\subsection{Discrete-Time Quantum Walks}

Consider a general multigraph $\mathcal{G}(V, E, f)$, according to definition 1. Associated to the vertices of $\mathcal{G}$, there exists a set of $|V|=n$ basis vectors $\{|v_j \rangle: v_j \in \mathbb{Z}\}$ that span an $n$-dimensional Hilbert space $H_P$, called the position space. Associated to the arcs of the graph, there exists a set of $m$ basis vectors $\{|c_i \rangle: c_i \in \mathbb{Z}\}$ that span an $m$-dimensional Hilbert space $H_C$, called the coin space. Each coin state is associated to a different subset of arcs in $\mathcal{G}$, e.g., in the DTQW on a line \cite{Venegas-Andraca2012, portugal13} coin states $|1\rangle$ and $|0\rangle$ are associated to the subsets of arcs pointing to the left and to the right, respectively. Thus, the quantum state of a walker, defined as

\begin{equation}
    |\psi(t)\rangle = \sum\limits_{i}\sum\limits_{j}b_{ij}|c_i\rangle \otimes |v_i\rangle
    \label{common_state_walker}
\end{equation}
stores information of the walker's position and which are the arcs of $\mathcal{G}$ through which the walker is allowed to move, given a certain position. 

Next, we can group the states that share the same coin state to obtain
\begin{equation}
    |\psi(t)\rangle = \sum\limits_{i}|c_i\rangle \otimes |V_i(t)\rangle
    \label{general_state_walker}
\end{equation}
where $|V_i(t) \rangle = \sum\limits_{j} b_{ij}|v_j\rangle$. Each subvector $|V_i(t) \rangle$ has a similar function than a probability vector in a random walk (Eq. (\ref{probability_vector})), with the difference that $|V_i(t) \rangle$ stores probability amplitudes. If the bases for $H_C$ and $H_P$ are the standard bases, then the $j$th entry of $|V_i(t) \rangle$ contains the probability amplitude corresponding to the walker with coin state $|c_i\rangle$ and position state $|v_j\rangle$. From a vector approach, $|\psi(t) \rangle$ can be described as a $n\cdot m$ column vector that is subdivided into $m$ vectors of size $n$. Explicitly

\begin{align}
\label{explicit_walker_state}
|\psi(t) \rangle =
\begin{pmatrix}
|V_0(t) \rangle \\
|V_1(t) \rangle \\
\vdots \\
|V_{m-1}(t) \rangle
\end{pmatrix}
\end{align}

Similar to the transition matrix in the random walk model, there exists an operator that contains information about the dynamics of the walker on the graph, with the difference that this operator is represented by a unitary matrix instead of a stochastic one. We call it the evolution operator of the system, and is defined as 

\begin{equation}
    \label{ev_operator_qw}
    U=S (C\otimes I_n)
\end{equation}
Where $C$ is the coin operator, and it acts on $|\psi(t) \rangle$ such that when $C\otimes I_n$ is applied, it affects only the coin register, and sets coin states into superposition. $S$ is the shift operator, and the objective when applied to $(C\otimes I_n)|\psi(t) \rangle$ is to make the position states transition from a state $|v_j\rangle$ to a state $|v_k \rangle$, which corresponds to a step to an adjacent node in $\mathcal{G}$. Notice that, in general, the shift operator might affect both the coin and position registers, but the coin register should not affect the position register, otherwise we would perform a quantum walk on two different graphs, similar to applying two different transition matrices to the same probability vector in a random walk.

Let $|\psi_0\rangle$ be the initial state of the system. The quantum walk starts when the operator $U$ is applied on $|\psi_0\rangle$. After $t$ steps the state of the system is given by

\begin{equation}
    |\psi(t)\rangle = U^t |\psi_0 \rangle
\end{equation}

In view of this description, we can conceive a DTQW as a superposition of walks that happen simultaneously on the different subgraphs of $\mathcal{G}$ associated to each coin state of the system.

Finally, we define the measurement operator of the state $|v_j\rangle$ as
\begin{align*}
M_k = \sum\limits_{i=0}^{m-1}|c_i\rangle \langle c_i| \otimes |v_k\rangle \langle v_k|=I_m \otimes |v_k\rangle \langle v_k|    
\end{align*}
The action of $M_k$ on the state of a quantum walker after $t$ steps, $|\psi(t)\rangle = \sum\limits_{i}|c_i\rangle \otimes |V_i(t)\rangle$, is to leave it in a superposition of all composite states with the same position state regardless of their coin state, that is

\begin{align*}
M_k|\psi(t)\rangle = \left(I_m\otimes|v_k\rangle\langle v_k|\right)\left(\sum\limits_{i}|c_i\rangle \otimes |V_i(t)\rangle\right)
\end{align*}

\begin{align*}
    M_k|\psi(t)\rangle=\sum\limits_i|c_i\rangle \otimes |v_k\rangle \hspace{9.6em}
\end{align*}
Thus, the probability of finding the state $|v_k \rangle$ after $t$ steps is given by

$P(|v_k\rangle) = \langle \psi(t)|M_k^{\dagger}M_k|\psi(t) \rangle$

\section{Adjacency Matrix Decomposition}

This section includes propositions and theorems obtained from the matrix study of the evolution operator of unitary coined discrete-time quantum walks. The first goal is to provide a mapping for a general graph $\mathcal{G}$ and its adjacency matrix into a directed multigraph $\mathcal{G}'$ and a unitary operator, respectively. The quantum walk will take place on $\mathcal{G}'$ and the unitary operator form of the adjacency matrix will serve as the shift operator of the system. The second goal is to study the transformation the shift operator and the graph $\mathcal{G}'$ undergo when then coin operator is applied. Finally, as a third goal we study the link between a general unitary operator --- and as a consequence a general quantum circuit --- and quantum walks.

\begin{figure}[h!]
\centering
\subfigure[]{
\resizebox{0.7\textwidth}{!}{ 
\includegraphics{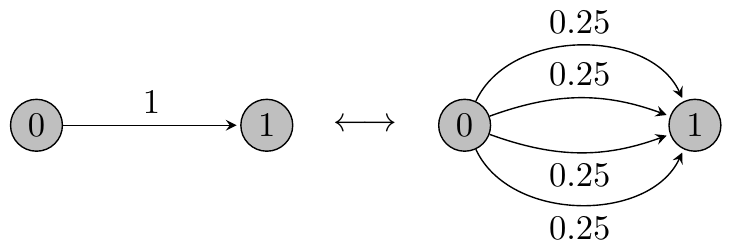}
\label{edge_transformations_a}
}
}
\subfigure[]{
\resizebox{0.7\textwidth}{!}{ 
\includegraphics{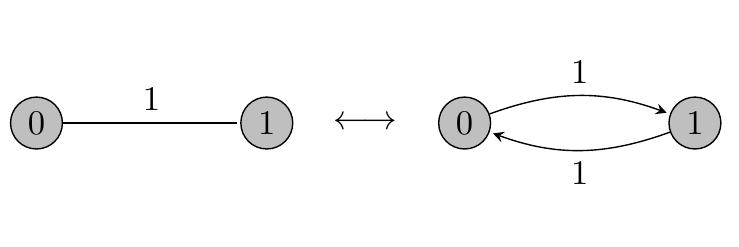}
\label{edge_transformations_b}
}
}
\caption{Examples of edge transformations. (a) Transformation between a directed edge and multiple directed edges pointing in the same direction. (b) Transformation between an undirected edge and two directed edges pointing in opposite directions.}
\label{edge_transformations}
\end{figure}

\noindent
\textbf{Proposition 1.}
Edges can be transformed in the following manner: 

\begin{enumerate}
    \item A directed edge $(v_i, v_j, e)$ with weight $w$ can be split into multiple directed edges $(v_i, v_j, e_k)$ with weights $w_k$ such that $w=\sum\limits_k w_k$.
    
    \item An undirected edge $\{v_i, v_j, e\}$ with weight $w$ can be split into two directed edges $(v_i, v_j, e_1)$ and $(v_j, v_i, e_2)$ each with weight $w$. Furthermore, the directed edges can be split into multiple edges using statement 1.
\end{enumerate}
For both statements, the converse is also true. Examples of each statement can be found in Fig. \ref{edge_transformations}.

The matrix representation of both transformations is the same, given that the $in-$ and $out-$ $degrees$ in both cases are preserved. 

\begin{figure}[h!]
\centering
\subfigure[]{
\centering
\resizebox{0.5\textwidth}{!}{ 
\includegraphics{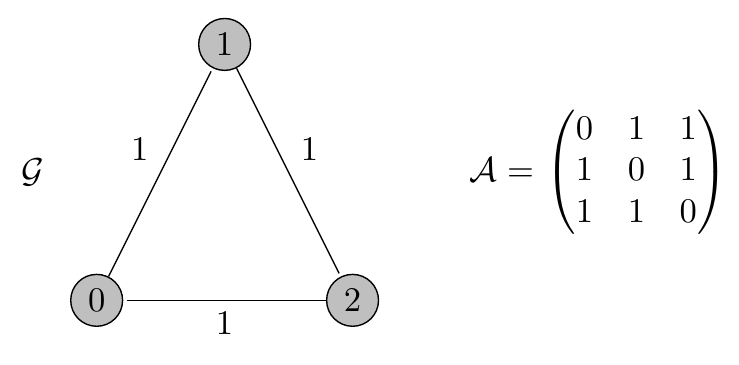}
}
}
\subfigure[]{
\centering
\resizebox{0.9\textwidth}{!}{    
\includegraphics{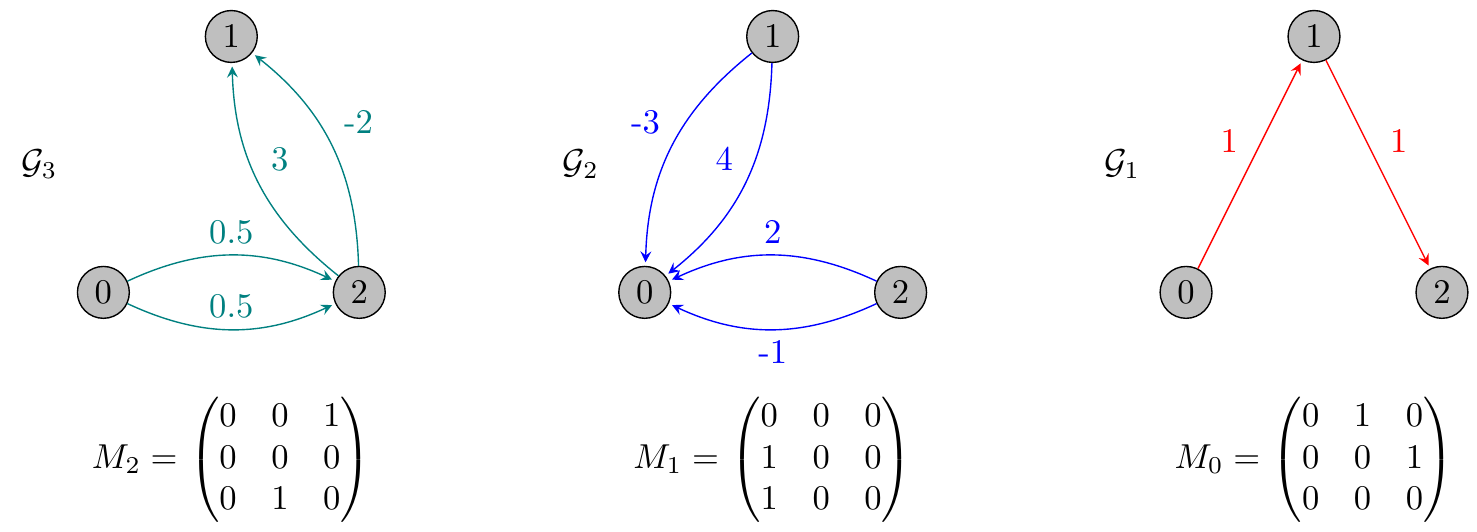}
}
}
\subfigure[]{
\centering
\resizebox{0.65\textwidth}{!}{ 
\includegraphics{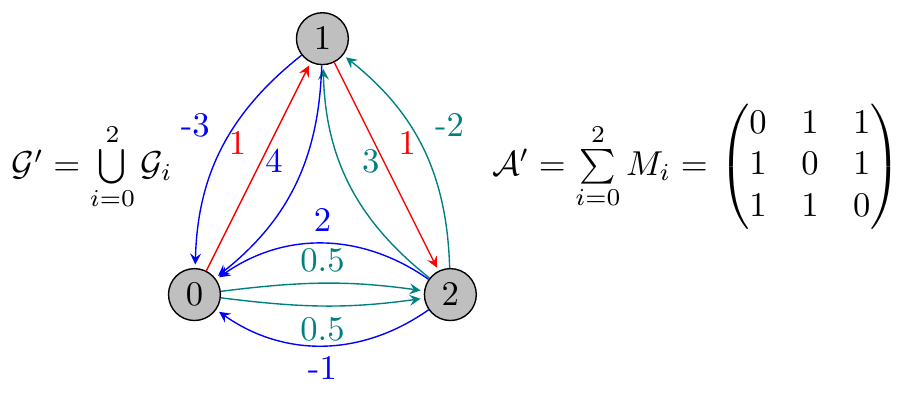}
}
}
\caption{Transformation of an undirected graph into a multigraph. (a) Displays the initial graph $\mathcal{G}(V, E, f)$ along with its adjacency matrix $\mathcal{A}$. (b) Displays three directed multigraphs $\mathcal{G}_i(V, E_i, f_i)$ with adjacency matrices $M_i$ such that $\mathcal{A}=\sum\limits_{i=0}^{2} M_i$. (c) Displays the union of the three multigraphs $\mathcal{G}_i$, as described in Proposition 2.}
\label{mapping_to_multigraph}
\end{figure}

\noindent
\textbf{Proposition 2.}
Any graph $\mathcal{G}$ associated to an adjacency matrix $\mathcal{A}$ can be mapped into a directed multigraph $\mathcal{G}'$ associated to the same adjacency matrix if and only if $\mathcal{A}$ can be expressed as a sum of $n$ matrices $M_i$. 

\noindent
\textbf{Proof.}

Let $\mathcal{A}$ be the adjacency matrix of a general graph $\mathcal{G}(V,E,f)$. Let $\mathcal{A}$ be decomposed into a sum of matrices, i.e. $\mathcal{A} = \sum\limits_{i=0}^{n-1} M_i$. Now let $\mathcal{G}_i(V,E_i,f_i)$ be a directed multigraph with adjacency matrix $M_i$, in such a way that all graphs $\mathcal{G}_i$ share the same vertex set, but not the same edge set $E_i$ and each edge set is associated to an incidence function $f_i$. Then, we define a new directed multigraph, $\mathcal{G}_i$, from the union of all the individual multigraphs, in the following way

\begin{equation}
\mathcal{G}'(V', E', f')=\bigcup\limits_{i=0}^{n-1}\mathcal{G}_i(V, E_i, f_i)    
\end{equation}
That is, $V'=V$, $E'=\bigcup\limits_{i=0}^{n-1}E_i$ and $f'=f_i$ $\forall i \in [0, n-1]$.

Fig. \ref{mapping_to_multigraph}, gives an example of how an undirected graph $\mathcal{G}$ can be transformed into a multigraph $\mathcal{G}'$ following this method.

Proposition 2 is equivalent to applying both statements of proposition 1 in $\mathcal{G}$ until no undirected edges are left. The new graph will be called $\mathcal{G}'$. $\mathcal{G}'$ will then be partitioned into $k$ subsets and each subset will be associated to an adjacency matrix. Graph $\mathcal{G}$ can be recovered by applying the inverse transformations to $\mathcal{G}'$.


\noindent
\textbf{Theorem 1.} 
The transpose of the adjacency matrix $\mathcal{A} \in \mathbb{C}^{n \times n}$ associated to a graph $\mathcal{G}$ of order $n > 1$ can be transformed into the shift operator $S$ of a quantum walk if and only if there exist $m^2$ operators ${\mathcal{B}^{\intercal}}_{ij}\in \mathbb{C}^{n \times n}$: $H_p \rightarrow H_p$ that satisfy the following equations simultaneously:

\begin{equation}
    \label{Andjecency_into_Kraus}
     \mathcal{A}^{\intercal} = \sum\limits_{i=0}^{m-1}\sum\limits_{j=0}^{m-1}{\mathcal{B}_{ij}^{\intercal}}
\end{equation}

\begin{equation}
    \label{kraus_completeness_relation_columns}
    \sum\limits_{i=0}^{m-1} {(\mathcal{B}^{\intercal}_{ij})^{\dagger}\mathcal{B}_{ik}^{\intercal}}=I_n \delta_{jk}
\end{equation}

\begin{equation}
    \label{kraus_completeness_relation_rows}
    \sum\limits_{i=0}^{m-1} {(\mathcal{B}^{\intercal}_{ji})^{\dagger}\mathcal{B}_{ki}^{\intercal}}=I_n \delta_{jk}
\end{equation}

\begin{equation}
\label{general_shift_tensor_kraus}
S=\sum\limits_{i=0}^{m-1}\sum\limits_{j=0}^{m-1} |c_i\rangle \langle c_j| \otimes {\mathcal{B}}_{ij}^{\intercal}
\end{equation}
where $\{|c_i\rangle \}$ is the canonical basis of $\mathbb{C}^m$.
Operators $\mathcal{B}_{ij}^{\intercal}$ are Kraus operators \cite{kraus_bohm_dollard_wootters_1983} and $m$ is the rank of the quantum channel they represent.
\noindent

\noindent
\textbf{Proof.}

Let $\mathcal{G}(V,E,f)$ be a graph of order $n > 1$ with adjacency matrix $\mathcal{A}$, where the quantum walk takes place. Let $|\psi\rangle = \sum\limits_i\sum\limits_k |c_i\rangle \otimes |v_k\rangle$ be the quantum state vector of a walker on $\mathcal{G}$, where the sets $\{|v_k\rangle: k\in \mathbb{Z}\}$ and $\{|c_i\rangle: i\in \mathbb{Z}\}$ are the canonical bases of $\mathbb{C}^n$ and $\mathbb{C}^m$, and are used as the bases for the Hilbert spaces $H_P$ and $H_C$, respectively. 

Suppose that the transpose adjacency matrix of $\mathcal{G}$ has additive decomposition \ $\mathcal{A}^{\intercal} = \sum\limits_{i=0}^{m-1}\sum\limits_{j=0}^{m-1} {\mathcal{B}^{\intercal}}_{ij}$, constrained to the completeness relation $\sum\limits_{i=0}^{m-1}{(\mathcal{B}^{\intercal})^{\dagger}_{ij}\mathcal{B}^{\intercal}_{ik}}=I_n \delta_{jk}$, which implies that the matrices of the sets $\{\mathcal{B}^{\intercal}_{ij}\}_{i=0}^{m-1}$, for fixed $j$, form sets of Kraus operators. Now, we decompose $\mathcal{G}(V,E,f)$ into subsets $\mathcal{G}_{ij}(V,E_{ij},f_i)$, each associated to an adjacency matrix $\mathcal{B}_{ij}$. That is, $\mathcal{G}_{ij}$ shares the same vertex set with the original graph $\mathcal{G}$, but holds only the arcs indicated by the adjacency matrix $\mathcal{B}_{ij}$. Then, we can associate a coin basis state $|c_j\rangle$ to all of the arcs of each subgraph $\mathcal{G}_j = \bigcup\limits_{i=0}^{m-1} \mathcal{G}_{ij}$, with adjacency matrix $\mathcal{A}_j=\sum\limits_{i=0}^{m-1}\mathcal{B}_{ij}$, in such a way that a walker with coin $|c_j\rangle$ can only move through arcs associated to that coin state. Thus, the most general shift operator that complies with this definition is the following

\begin{align*}
    S=\sum\limits_{i=0}^{m-1}\sum\limits_{j=0}^{m-1} |c_i\rangle \langle c_j| \otimes \mathcal{B}^{\intercal}_{ij},
\end{align*}
which explicitly is a block matrix of dimensions $nm \times nm$, i.e.

\begin{equation}
S=
\label{block_matrix_shift}
\begin{pmatrix}
\mathcal{B}^{\intercal}_{00} & \mathcal{B}^{\intercal}_{01} & \dots & \mathcal{B}^{\intercal}_{0  m-1} \\
\mathcal{B}^{\intercal}_{10} & \mathcal{B}^{\intercal}_{11} & \dots & \mathcal{B}^{\intercal}_{1  m-1} \\
\vdots & \vdots & \ddots & \vdots \\
\mathcal{B}^{\intercal}_{m-1  0} & \mathcal{B}^{\intercal}_{m-1  1} & \dots & \mathcal{B}^{\intercal}_{m-1  m-1}
\end{pmatrix}
\end{equation}
Furthermore, notice that due to the completeness relation in Eq. (\ref{kraus_completeness_relation_columns}), each of the sets of Kraus operators mentioned before contained in each column of $S$, which implies that $S$ is unitary. To prove the unitarity of $S$ consider 
{\small
\begin{align*}
    S^{\dagger}S = \left(\sum\limits_{k=0}^{m-1}\sum\limits_{l=0}^{m-1} |c_l\rangle\langle c_k| \otimes (\mathcal{B}^{\intercal}_{kl})^{\dagger}\right) 
    \left(\sum\limits_{r=0}^{m-1}\sum\limits_{s=0}^{m-1}|c_r\rangle \langle c_s| \otimes \mathcal{B}^{\intercal}_{rs}\right)
\end{align*}}
Which simplifies to
\begin{align*}
   S^{\dagger}S = \sum\limits_{l=0}^{m-1}\sum\limits_{s=0}^{m-1} |c_l\rangle \langle c_s| \otimes \sum\limits_{k=0}^{m-1}(\mathcal{B}_{kl}^{\intercal})^{\dagger} (\mathcal{B}^{\intercal}_{ks})
\end{align*}

Using the fact that $\sum\limits_{k=0}^{m-1} {(\mathcal{B}^{\intercal})^{\dagger}_{kl}\mathcal{B}^{\intercal}_{ks}}=I_n \delta_{ls}$, it follows that

\begin{align*}
    S^{\dagger}S = \sum\limits_{l=0}^{m-1}\sum\limits_{s=0}^{m-1} |c_l\rangle \langle c_s| \otimes I_n \delta_{ls}
\end{align*}

\begin{align*}
    = \sum\limits_{s=0}^{m-1} |c_s\rangle \langle c_s| \otimes I_n \hspace{-1.07em}
\end{align*}
Finally, because the set $\{|c_s\rangle\}$ is complete, $\sum\limits_{s=0}^{m-1} |c_s\rangle \langle c_s|=I_m$. Thus,
\begin{align*}
    S^{\dagger}S=I_m\otimes I_n = I_{nm}.
\end{align*}

The transpose of a unitary is also unitary, thus repeating the same process for $S^{\intercal}$, we get to the conclusion that eq. (\ref{kraus_completeness_relation_rows}) is necessary. That is, the rows of the shift operator also form sets of Kraus operators each.

To study the action of $S$, consider the general state of a quantum walker, given by Eq. (\ref{general_state_walker}). Applying $S$ to $|\psi(t) \rangle$, we obtain

{\small
\begin{align*}
    S|\psi \rangle = \left(\sum\limits_{i=0}^{m-1}\sum\limits_{j=0}^{m-1} |c_i\rangle \langle c_j| \otimes \mathcal{B}^{\intercal}_{ij}\right)
    \left(\sum\limits_{s}|c_s\rangle \otimes |V_s(t)\rangle\right)
\end{align*}
}
Which simplifies to 
\begin{align*}
    S|\psi \rangle = \sum\limits_{i} |c_i\rangle \otimes \sum\limits_{j}\mathcal{B}^{\intercal}_{ij}|V_i(t)\rangle 
\end{align*}
Given that $\mathcal{B}^{\intercal}_{ij}: H_p \rightarrow H_p$, the operation $\mathcal{B}^{\intercal}_{ij}|V_i(t)\rangle$ yields to a new superposition of position states, $\mathcal{B}_{ij}|V_i(t)\rangle = \sum\limits_{k} b'_{ik}|v_k\rangle$. The interpretation is that Kraus operators, $\mathcal{B}_{ij}$, allow walkers to shift between adjacent vertices of the subgraphs $\mathcal{G}_{ij}$ they are associated with, similar to the action of the transition matrix in a random walk (see Eq. \eqref{recurrence_random_walk}).

In view that $S$ is a unitary operator composed of Kraus operators that act on the position register, we conclude that $S$ is suitable to be the shift operator of a quantum walk, which proves the conditional statement.

For the converse statement, suppose $S$ is a unitary matrix of size $nm \times nm$, with $n,m > 1$. Because of its dimensions, $S$ can be subdivided in $m^2$ blocks matrices of size $n$ each, and be written as

\begin{equation}
    S=\sum\limits_{i=0}^{m-1}\sum\limits_{j=0}^{m-1} |c_i\rangle \langle c_j| \otimes \mathcal{B}^{\intercal}_{ij}
\end{equation}

\noindent
where $|c_i\rangle$ and $|c_j\rangle$ are vectors of the canonical basis in $\mathbb{C}^m$. 

The unitarity condition of $S$, implies Eq. (\ref{kraus_completeness_relation_columns}). Considering
{\small
\begin{align*}
    S^{\dagger}S=\left(\sum\limits_{i=0}^{m-1}\sum\limits_{j=0}^{m-1} |c_j\rangle \langle c_i| \otimes (\mathcal{B}^{\intercal}_{ij})^{\dagger}\right)\left(\sum\limits_{k=0}^{m-1}\sum\limits_{l=0}^{m-1} |c_k\rangle \langle c_l| \otimes \mathcal{B}^{\intercal}_{kl}\right)
\end{align*}
}

\begin{align*}
    =\sum\limits_{j=0}^{m-1}\sum\limits_{l=0}^{m-1} |c_j\rangle \langle c_l| \otimes \sum\limits_{i=0}^{m-1}(\mathcal{B}^{\intercal}_{ij})^{\dagger} \mathcal{B}^{\intercal}_{il}=I_{nm} \hspace{2.1em}
\end{align*}
This equation can only be true if 

\begin{align*}
    \sum\limits_{i=0}^{m-1}{(\mathcal{B}^{\intercal})^{\dagger}_{ij}\mathcal{B}^{\intercal}_{il}}=I_n \delta_{jl} 
\end{align*}
which means that the set of matrices $\mathcal{B}^{\intercal}_{ij}$ of each column in $S$ forms a set of Kraus operators. Again, the same process can be followed for $S^{\intercal}$ to obtain Eq. (\ref{kraus_completeness_relation_rows}).

Given that the operators $\mathcal{B}_{ij}$ are square matrices of the same dimension, each of them can be associated to a graph $\mathcal{G}_{ij}$. The graph of the system is then defined by $\mathcal{G}=\mathcal{G}(V, \bigcup\limits_{i=0}^{m-1}\bigcup\limits_{j=0}^{m-1}E_{ij}, f)$, i.e. $\mathcal{G}$ is composed of the union of arc sets of all the subgraphs $\mathcal{G}_{ij}$, keeping the same vertex set for all of them, as described in Proposition 2. The adjacency matrix of $\mathcal{G}$ is then given by 
\begin{equation}
    \mathcal{A} = \sum\limits_{i=0}^{m-1}\sum\limits_{j=0}^{m-1} \mathcal{B}_{ij}.
\end{equation}

The transformation of the converse statement is not unique, given that we can vary the dimensions of the basis vectors $|c_i\rangle$ and block matrices $\mathcal{B}^{\intercal}_{ij}$, with the only condition that the dimension of $S$ remains fixed, that is the new dimensions $n'$ and $m'$ of the position and coin spaces must comply with $n'm'=nm$. This completes the proof.

A consequence of the additive decomposition method is that it might change the distribution of weights in the original graph and even induce new connections between vertices in order to comply with the unitarity conditions, e.g. two originally non-connected vertices might be connected after being mapped into two parallel arcs pointing in the same direction with positive and negative weights of the same magnitude. 

Next we present a corollary that follows from Theorem 1.

\noindent
\textbf{Corollary 1.}

If $\mathcal{A}^{\intercal}$ can be decomposed as the sum of $m$ unitary matrices $\mathcal{B}^{\intercal}_i$, the shift operator of the system is given by

\begin{equation}
S=\sum\limits_{i=0}^{m-1} |c_i\rangle \langle c_i| \otimes {\mathcal{B}}^{\intercal}_{i},
\end{equation}
or explicitly

\begin{equation}
S =
\label{block_diag_shift}
\begin{pmatrix}
\mathcal{B}^{\intercal}_0 & 0 & \dots & 0 \\
0 & \mathcal{B}^{\intercal}_1 & \dots & 0 \\
\vdots & \vdots & \ddots & \vdots \\
0 & 0 & \dots & \mathcal{B}^{\intercal}_{m-1} 
\end{pmatrix}
\end{equation}

\noindent
\textbf{Proof.}

This is the special case of Theorem 1 in which the sets of Kraus operators $\mathcal{B}^{\intercal}_{ij}$ consist of a unitary matrix for equal indices, and a zero matrix for different indices, i.e. $\mathcal{B}^{\intercal}_i = \mathcal{B}^{\intercal}_{ij} \delta_{ij}$. In other words, the Kraus rank for all the sets of Kraus operators is 1.


This corollary  is specially useful for the quantum circuit implementation of shift operators, and was studied by Montanaro \cite{montanaro_2007} and Zhan \cite{godsil_zhan_2019} for the particular case where the Kraus operators are permutation matrices.

Now that we have analyzed in detail the matrix form of the shift operator of a quantum walk, the link between this operator and quantum circuits emerges as a corollary, which is presented next.

\noindent
\newline{}
\textbf{Corollary 2.}
Any quantum circuit acting on a bipartite system is suitable to be the shift operator in a coined quantum walk.
\noindent

\noindent
\textbf{Proof.}

Any unitary matrix has a quantum gate representation. A quantum circuit is no more than a set of consecutive applications of quantum gates, which is in turn unitary. Thus, we can always apply Theorem 1 to any quantum circuit that acts on a bipartite quantum register, the circuit to a multigraph to perform a DTQW.



\begin{figure}[h!]
\centering

\begin{minipage}[t]{.4\textwidth}
\subfigure[]{
\resizebox{\textwidth}{!}{    
\includegraphics{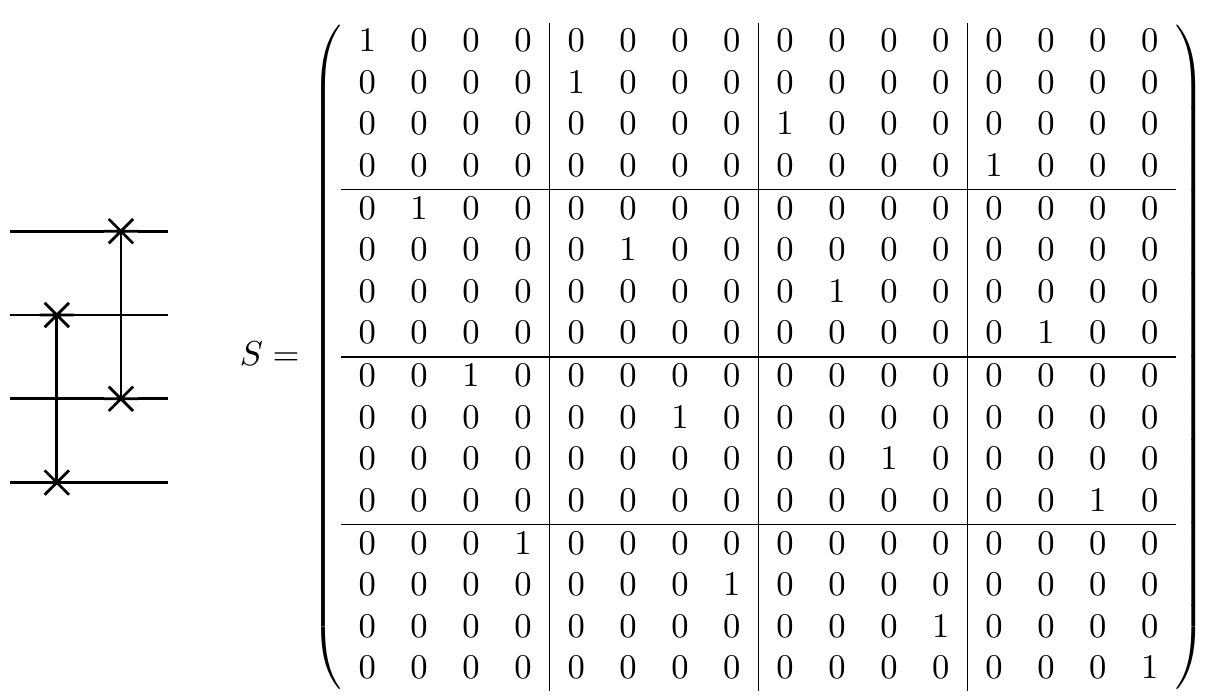}
\label{model_circuit_example_a}
}
}

\subfigure[]{
\resizebox{\textwidth}{!}{    
\includegraphics{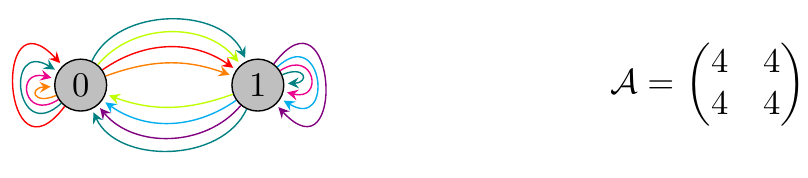}
\label{model_circuit_example_b}
}
}

\subfigure[]{
\resizebox{\textwidth}{!}{    
\includegraphics{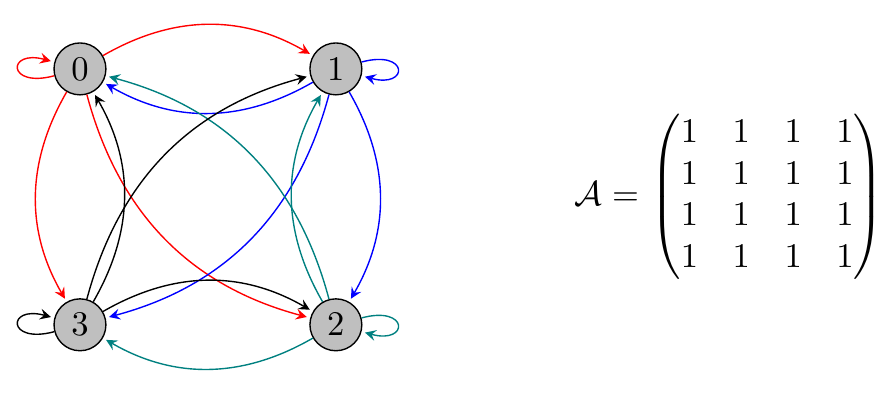}
\label{model_circuit_example_c}
}
}

\subfigure[]{
\resizebox{\textwidth}{!}{    
\includegraphics{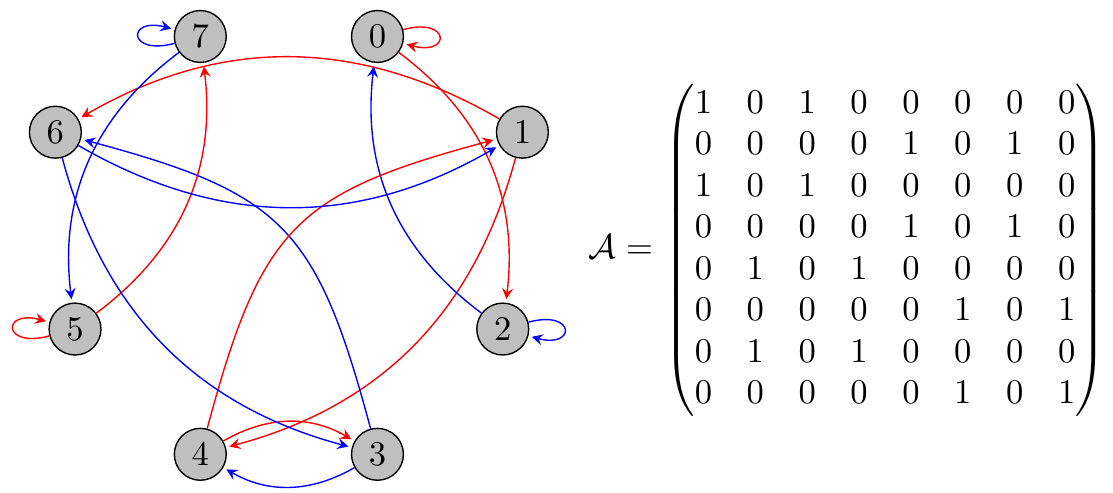}
\label{model_circuit_example_d}
}
}
\end{minipage}
\begin{minipage}[t]{.4\textwidth}
\subfigure[]{
\resizebox{\textwidth}{!}{    
\includegraphics{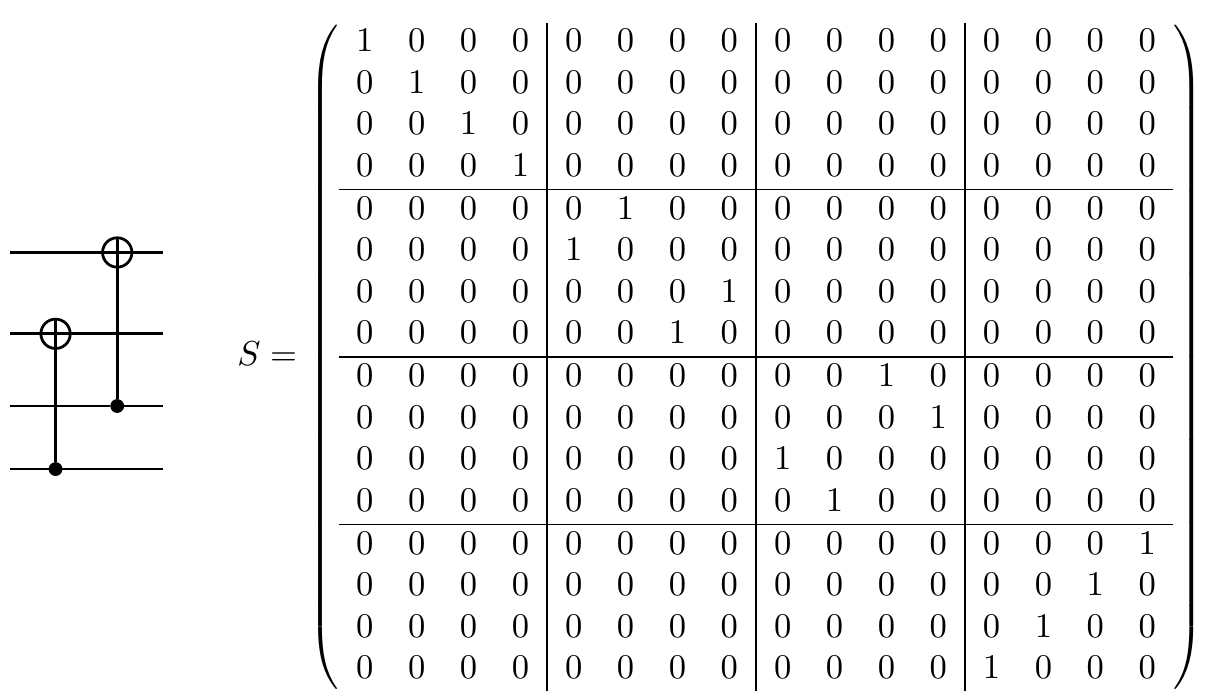}
\label{model_circuit_corollary_example_a}
}
}

\subfigure[]{
\resizebox{\textwidth}{!}{    
\includegraphics{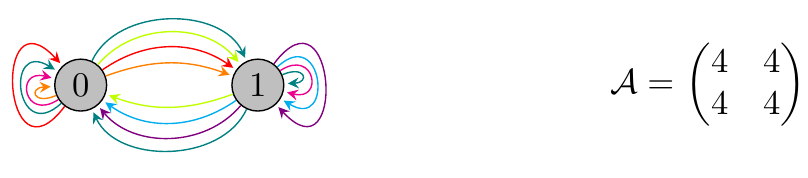}
\label{model_circuit_corollary_example_b}
}
}

\subfigure[]{
\resizebox{\textwidth}{!}{    
\includegraphics{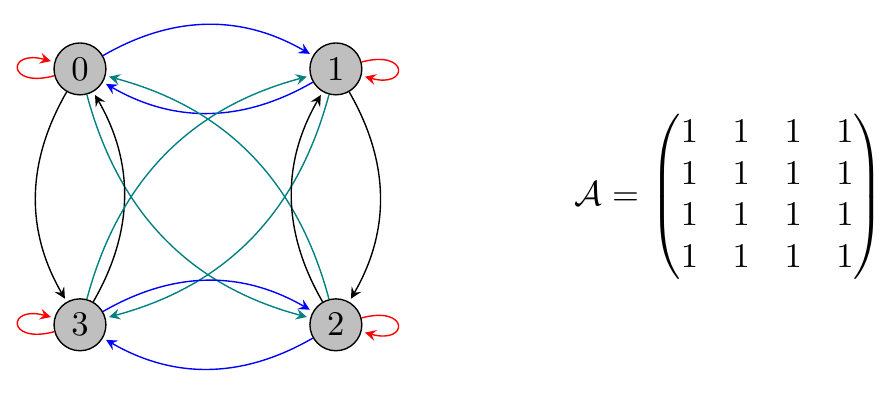}
\label{model_circuit_corollary_example_c}
}
}

\subfigure[]{
\resizebox{\textwidth}{!}{    
\includegraphics{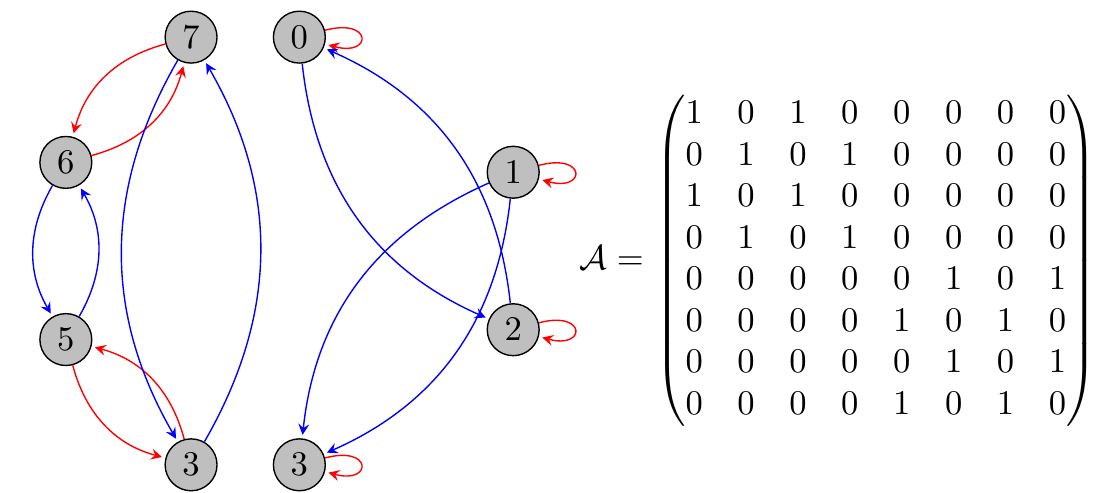}
\label{model_circuit_corollary_example_d}
}
}
\end{minipage}

\caption{(a) Quantum circuit along with its unitary matrix representation. The sizes of the coin $(m)$ and position $(n)$ registers can vary, holding $n+m=4$. This partitions $S$ into $2^m$ matrices of sizes $2^n$. $S$ is subdivided for the case $m=2, n=2$. (b), (c) and (d) are multigraphs and adjacency matrices associated to $S$ when $m=1$, $n=3$; $m=2$, $n=2$ and $m=3$, $n=1$, respectively. Colors are used to distinguish the subgraphs associated to different sets of Kraus operators. Red, blue, green and black arcs in (c) are associated to the first, second, third and forth columns in the example operator $S$ in (a). Each color is associated to a different coin state. Notice that when $m=2$ and $n=2$ the quantum circuit corresponds to the shift operator a quantum walk on a complete graph of four nodes with self-loops as proposed in \cite{douglas_wang_2009}.}
\label{model_circuit_example}
\end{figure}

Fig. \ref{model_circuit_example} presents two examples that together summarize Theorem 1 and both corollaries. In Fig. \ref{model_circuit_example_a}, we display the circuit proposed by Douglas and Wang \cite{douglas_wang_2009} for a DTQW on a complete graph of four vertices, which uses four qubits. Next to the circuit we also display the unitary matrix, $S$, associated to it. If we consider the total quantum register to be composed of two subregisters, then $S$ will be automatically split into block matrices. Consider the case where the two bottom qubits correspond to the coin register and the two upper ones to the position register, then $S$ will be a $4 \times 4$ block matrix where each block will be of size $4 \times 4$. In Fig. \ref{model_circuit_example_a} the lines in  $S$ make explicit this partition, and Fig. \ref{model_circuit_example_c} displays the graph and adjacency matrix associated to $S$, and in consequence to the circuit. Fig. \ref{model_circuit_example_b} corresponds to the mapping when the coin and position registers consist of 1 and 3 qubits, and \ref{model_circuit_example_d} corresponds to the mapping when the coin and position registers consist of 3 and 1 qubits, respectively.


In Fig. \ref{model_circuit_corollary_example_a} we present an alternative version of the circuit for a DTQW on a complete graph of four vertices, but in this case the associated unitary matrix $S$ consists of a block diagonal matrix, thus perfectly exemplifies corollary 1. Likewise, if we consider the coin and position registers to be both of size 2, then we obtain the partition indicated by the lines in $S$. Fig. \ref{model_circuit_corollary_example_c} is the associated mapping of this partition. In the case where the coin and position registers have sizes 1 and 3, we obtain the mapping of Fig. \ref{model_circuit_corollary_example_b}, and when the coin and position registers have sizes 3 and 1, we obtain Fig. \ref{model_circuit_corollary_example_d}, respectively.

\newpage{}
\noindent
\textbf{Proposition 3.} 


Given a general shift operator $S$, associated to a graph $\mathcal{G}$ as described in Theorem 1, the evolution operator of a coined quantum walk can be completed by choosing a general coin of the form

\begin{equation}
    \label{general_coin}
    \mathcal{C}=\sum\limits_{k=0}^{n-1} C_{k} \otimes |v_k\rangle \langle v_k|
\end{equation}
where all operators $C_k$ are unitary of size $m \times m$. 

Furthermore, if all operators $C_k$ are the same, we obtain the usual form of the coin operator, i.e.
\begin{equation}
    \mathcal{C} = C \otimes I
\end{equation}


\noindent
\textbf{Proof.}

It is convenient to start the analysis with the simplest case to comprehend the general idea. Let $C$ be an arbitrary unitary matrix, then the operator $C\otimes I_n$ takes the following form in explicit matrix notation

\begin{equation}
C \otimes I_n=
\begin{pmatrix}
c_{00}I_n & c_{01}I_n & \dots & c_{0 m-1} I_n \\
c_{10}I_n & c_{11}I_n & \dots & c_{1 m-1} I_n \\
\vdots & \vdots & \ddots & \vdots \\
c_{m-10}I_n & c_{m-11}I_n & \dots & c_{m-1m-1}I_n
\end{pmatrix}
\end{equation}
where weights $c_{ij}$ are the entries of $C$.

Now let the shift operator of a quantum walk take the form of Eq. (\ref{block_matrix_shift}). Then, the evolution operator can be written in block matrix notation as 


{\small 
\begin{multline}
U=S(C\otimes I_n)=
\label{evolution_op_block_kraus}
\addtolength{\arraycolsep}{-3pt}
\hfilneg
\begin{pmatrix}
\sum\limits_{i=0}^{m-1}c_{i0}\mathcal{B}^{\intercal}_{0i} & \sum\limits_{i=0}^{m-1}c_{i1}\mathcal{B}^{\intercal}_{0i} & \dots & \sum\limits_{i=0}^{m-1}c_{im-1}\mathcal{B}^{\intercal}_{0i} \\
\sum\limits_{i=0}^{m-1}c_{i0}\mathcal{B}^{\intercal}_{1i} & \sum\limits_{i=0}^{m-1}c_{i1}\mathcal{B}^{\intercal}_{1i} & \dots & \sum\limits_{i=0}^{m-1}c_{im-1}\mathcal{B}^{\intercal}_{1i} \\
\vdots & \vdots & \ddots & \vdots \\
\sum\limits_{i=0}^{m-1}c_{i0}\mathcal{B}^{\intercal}_{m-1i} & \sum\limits_{i=0}^{m-1}c_{i1}\mathcal{B}^{\intercal}_{m-1i} & \dots & \sum\limits_{i=0}^{m-1}c_{im-1}\mathcal{B}^{\intercal}_{m-1i}
\end{pmatrix}
\hspace{1000 pt minus 1fil}
\end{multline}
}

Similar to the case of the shift operator described in Theorem 1, the complete evolution operator has also a block matrix form, whose columns constitute sets of Kraus operators, and walkers with coin state $|c_j\rangle$ move through the subgraph, $\mathcal{G}_j$, whose transpose adjacency matrix, $\mathcal{A}^{\intercal}_j$, is obtained by the addition of all block elements of column $j$. This becomes evident when considering the explicit vector form of the state of a walker (eq. \ref{explicit_walker_state}), from which we can see that in $U|\psi(t)\rangle$, the elements of column $j$ act on the substate $|V_j(t)\rangle$. The transpose adjacency matrix of $\mathcal{G}_j$ has the following form

\begin{align}
    \mathcal{A}^{\intercal}_j = \sum_{k=0}^{m-1}\sum_{i=0}^{m-1}c_{kj}\mathcal{B}^{\intercal}_{ik}
    \label{weigthed_kraus_adj}
\end{align}
That is, each time the $U$ operator is applied on $|\psi(t)\rangle$, all the position states with coin state $|c_j\rangle$ will be modified by the action of $\mathcal{A}^{\intercal}_j$, which is a weighted sum of all the Kraus operators that compose the original matrix $\mathcal{A}^{\intercal}$ (see Eq. \ref{Andjecency_into_Kraus}).

For the case of the general coin, suppose the $k$th coin $C_k$ has entries $c^k_{ij}$. Then, eq. (\ref{general_coin}) can also be written as a block matrix

\begin{equation}
\mathcal{C}=
\begin{pmatrix}
D_{00} & D_{01} & \dots & D_{0 m-1}  \\
D_{10} & D_{11} & \dots & D_{1 m-1} \\
\vdots & \vdots & \ddots & \vdots \\
D_{m-10} & D_{m-11} & \dots & D_{m-1m-1}
\end{pmatrix}
\end{equation}

\noindent
Where 
\begin{equation}
D_{ij}=
\begin{pmatrix}
c^0_{ij} & 0 & \dots & 0 \\
0 & c^1_{ij} & \dots & 0 \\
\vdots & \vdots & \ddots & \vdots \\
0 & 0 & \dots & c^{m-1}_{ij} 
\end{pmatrix}
\end{equation}
Thus, 

{\small 
\begin{multline}
U=S\mathcal{C}=
\label{evolution_op_general_coin}
\addtolength{\arraycolsep}{-3pt}
\hfilneg
\begin{pmatrix}
\sum\limits_{i=0}^{m-1}\mathcal{B}^{\intercal}_{0i}D_{i0} & \sum\limits_{i=0}^{m-1}\mathcal{B}^{\intercal}_{0i}D_{i1} & \dots & \sum\limits_{i=0}^{m-1}\mathcal{B}^{\intercal}_{0i}D_{im-1} \\
\sum\limits_{i=0}^{m-1}\mathcal{B}^{\intercal}_{1i}D_{i0} & \sum\limits_{i=0}^{m-1}\mathcal{B}^{\intercal}_{1i}D_{i1} & \dots & \sum\limits_{i=0}^{m-1}\mathcal{B}^{\intercal}_{1i}D_{im-1} \\
\vdots & \vdots & \ddots & \vdots \\
\sum\limits_{i=0}^{m-1}\mathcal{B}^{\intercal}_{m-1i}D_{i0} & \sum\limits_{i=0}^{m-1}\mathcal{B}^{\intercal}_{m-1i}D_{i1} & \dots & \sum\limits_{i=0}^{m-1}\mathcal{B}^{\intercal}_{m-1i}D_{im-1}
\end{pmatrix}
\hspace{1000 pt minus 1fil}
\end{multline}
}

\begin{figure}[b!]
\centering
\subfigure[]{
\resizebox{0.45\textwidth}{!}{    
\includegraphics{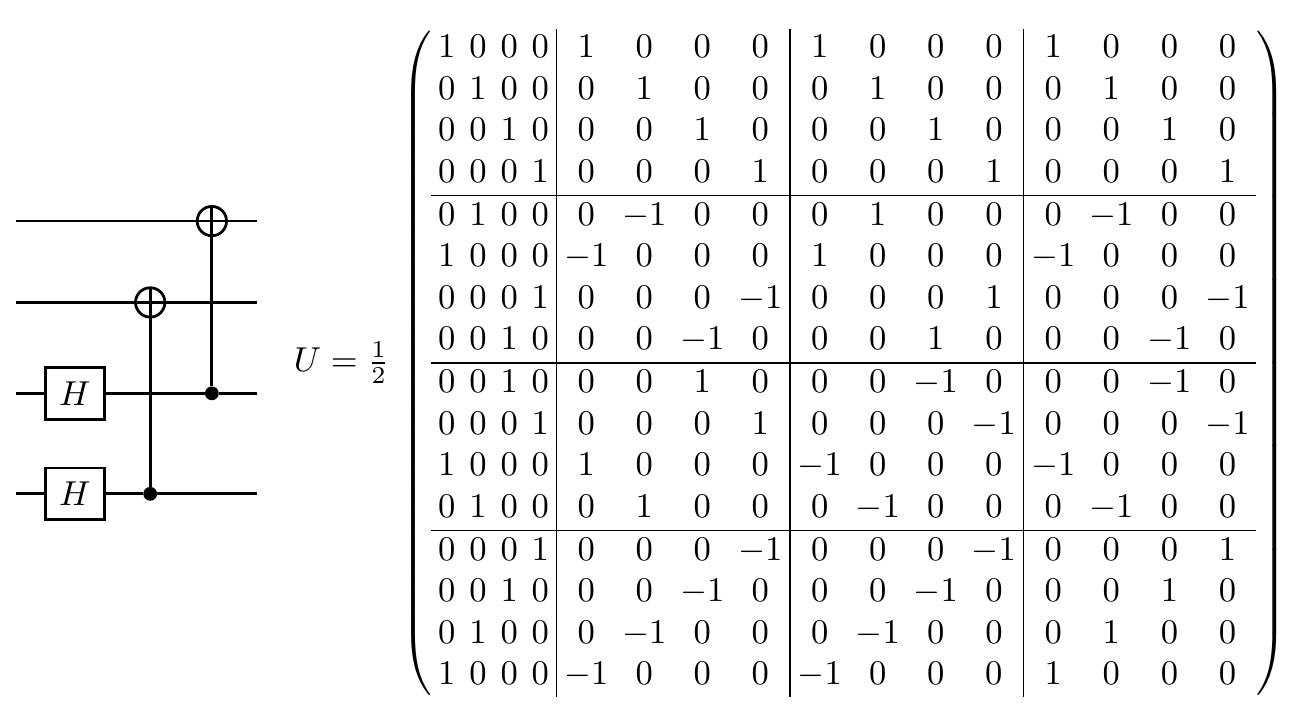}
}
}
\subfigure[]{
\resizebox{0.45\textwidth}{!}{    
\includegraphics{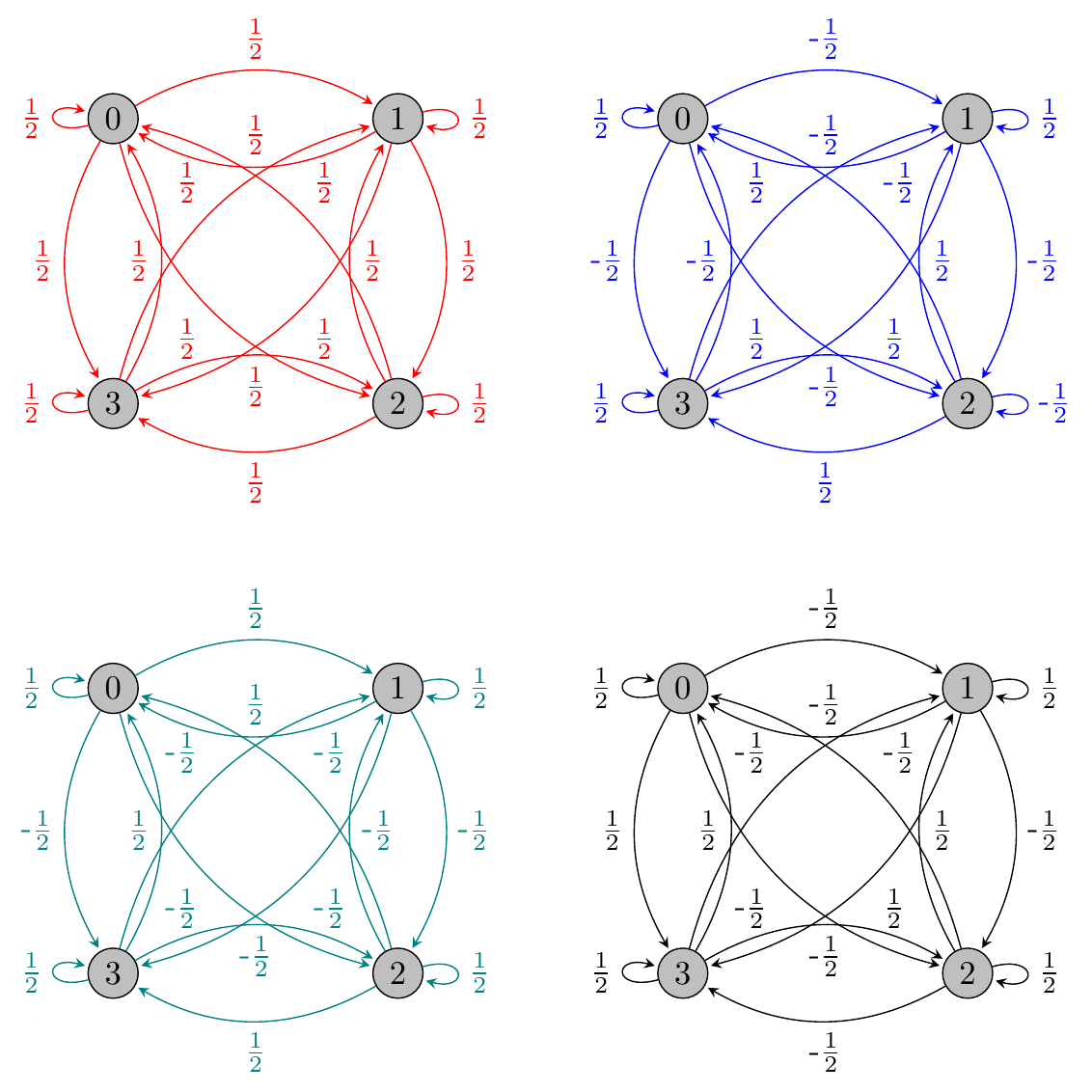}
}
}
\caption{(a) {\small Displays the full evolution operator in both circuit and matrix form of a quantum walk on a complete graph of four nodes with self-loops, for the case where the position and coin register have sizes of $m=2$ and $n=2$, respectively. (b) Displays the subgraphs associated to each of the columns of the operator $U$. The top-left, top-right, bottom-left and bottom-right graphs correspond to the first, second, third and fourth columns of $U$, respectively. The graph associated to $U$ is given by the union of the edges of the four subgraphs, using the same vertex set for all of them, as shown in Fig. \ref{mapping_to_multigraph}. Notice that the subgraph associated to each column of $U$ is similar to the total graph associated to the shift operator $S$ in Fig. \ref{model_circuit_corollary_example_c}, changing only the weights of the arrows according to the entries of the matrix representation of the coin operator.}}
\label{simultaneos_walks_swap}
\end{figure}

The operation $\mathcal{B}^{\intercal}_{ij}D_{ji}$ scales the whole $l$th column of $\mathcal{B}^{\intercal}_{ij}$ by the factor $c^l_{ji}\in D_{ji}$ and, the $l$th column of $\mathcal{B}^{\intercal}$ contains the weights of all the arcs that come out of the $l$th node towards other nodes. 

Similar to the former case, the walker with coin state $|c_j\rangle$ evolves according to the matrix
\begin{align}
    \mathcal{A}^{\intercal}_j = \sum_{k=0}^{m-1}\sum_{i=0}^{m-1}\mathcal{B}^{\intercal}_{ik} D_{kj}
\end{align}

In principle, there are no restrictions on the values the scaling factors can take in both cases studied if the coin operator is unitary. Thus we conclude that any unitary matrix applied on the coin register is suitable to complete the evolution operator of the quantum walk. 

\newpage{} 
The action of applying a coin operator to a shift operator is to split each of the arcs of the of the graph associated to $S$ into $2^m$ new arcs, one associated to a different coin state. If we consider the subsets of edges associated to the same coin state, we find that each subset forms a graph with a similar structure as the one associated to the shift operator. In other words, this can be interpreted as the action of the evolution operator is to create a superposition of $2^m$ graphs with the same connections as the one associated to $S$ but modifying its weights. 

To exemplify this, consider the quantum circuit of Fig. \ref{model_circuit_corollary_example_a}, if we take the first two qubits of the register to represent the position state and the last two qubits to represent the coin state, the associated graph to this system is given by Fig. \ref{model_circuit_corollary_example_c}. Now see from Fig. \ref{simultaneos_walks_swap}, that if we apply a Hadamard coin to the circuit shown in \ref{model_circuit_corollary_example_a} to complete the evolution operator, the associated graph will be a superposition of $4$ like graphs, with the difference that the weights of the new graphs will be modified, according to the entries of the coin operator matrix. Notice that as a consequence, given that the coin operator modifies the shift operator, if Theorem 1 is applied to construct $S$, the sum of all the Kraus operators in $U$ will not yield back the original adjacency matrix given that the coin operator modifies the weights in $S$.

\begin{figure}[h!]
\centering
\subfigure{\includegraphics{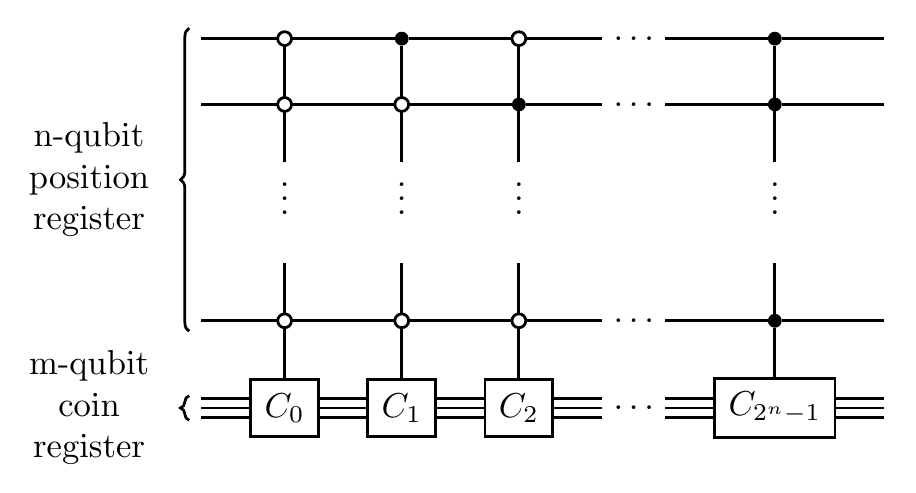}}
\subfigure{\includegraphics{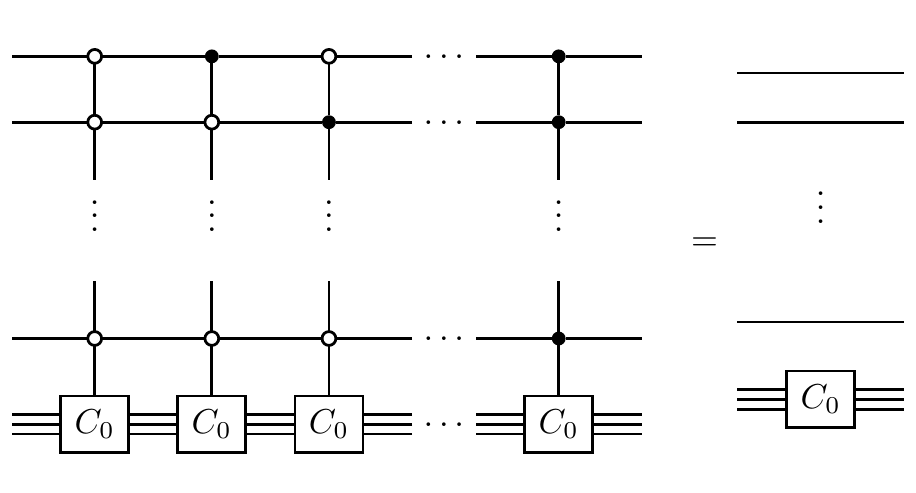}}
\caption{(a) Consider black dots to be 1's and white dots to be 0's. The sequence of a general coin is completed by controlling $2^n$ gates using all binary strings. If less than $2^n$ $C_i$ gates are fully controlled to generate a coin, the remaining ones will be the identity by default. (b) In the case in which all the multi-control gates have the same target gate $C_0$, the circuit can be reduced to the application of one single gate $C_0$ to the coin register.}
\label{general_coin_ciq}
\end{figure}

The circuit form of a general coin operator is displayed in Fig. \ref{general_coin_ciq}. In this circuit, we consider the white dots as 0's and the black dots as 1's, in such a way that we can see the sequence of white and black dots as binary code, where the less significant bit is given by the uppermost dot. Each operator $C_i$ acts only on the position state $|v_k\rangle$ whose index $k$ matches the binary representation of the sequence of white and black dots.

\section{Conclusion}\label{conclusion_section}

In this paper we have studied studied unitary coined DTQW from a matrix approach, from which we have been able to prove that the shift operator of a DTQW can be thought of as a quantum mechanical version of the transition matrix of a random walk. We did this by proving a set of necessary and sufficient equations with which we can map the transpose adjacency matrix associated to a graph $\mathcal{G}$ into the unitary operator of a bipartite quantum system, which we call shift operator. The shift operator can be applied on a quantum state of the bipartite system, modifying the probability amplitudes of the state, and, thus, generating a quantum walk. The bipartite state of a quantum walk can then be conceived as a scaled up version of a stochastic vector, which holds the information of multiple walks happening simultaneously on a multigraph version of $\mathcal{G}$, called $\mathcal{G}'$. A general description of the mapping from $\mathcal{G}$ to $\mathcal{G}'$ and of the dynamics of a quantum walker on $\mathcal{G}'$ was also given. 

The importance of the set of equations (\ref{Andjecency_into_Kraus})-(\ref{general_shift_tensor_kraus}) resides in the fact that, in general, it is difficult to obtain the evolution operator of a quantum walk for a specific graph. Now, given the adjacency matrix, $\mathcal{A}$, of the graph on which we want to perform the quantum walk, the process for mapping $\mathcal{A}^{\intercal}$ into $\mathcal{S}$ is standardized and can be automated. 

To complete the evolution operator, we also extended the idea of a quantum coin, in such a way that any controlled or non-controlled unitary operator can be suitable to be the coin operator of a DTQW. Although this fact might result in a biased DTQW in some cases, we do not see it as a reason to impose further restrictions on the coin operator. With this extension we can even choose a different coin for each vertex in the most extreme case, or make combinations of coins. That will depend on the purpose of the algorithm we want to run.

This way, we make progress on both of the problems stated in the first section of the paper, i.e. we present present a method to build the evolution operator of a unitary coined quantum walk from scratch, and describe the link between bipartite quantum circuits and quantum walks. Nevertheless, the set of graphs for which this work is applicable is only the one whose adjacency matrices can be decomposed as presented in Eq. (\ref{Andjecency_into_Kraus}). Regarding the link between quantum circuits and evolution operators of quantum circuits, we only provided a way to go from quantum circuits to evolution operators but not the converse, although the converse problem reduces to mapping a unitary matrix into a set of quantum gates, which is a problem that has been studied in different works \cite{krol_2019, unitary_decomposition_li}.


\section*{Acknowledgments}

Both authors acknowledge the financial support provided by Tecnologico de Monterrey, Escuela de Ingenieria y Ciencias and Consejo Nacional de Ciencia y Tecnología (CONACyT). SEVA acknowledges the support of CONACyT-SNI [SNI number 41594]. SEVA acknowledges the unconditional support of his family.

\section*{Data availability}
Data sharing not applicable to this article as no datasets were generated or analysed during the current study.

\section*{Conflict of Interest}
All authors declare that they have no conflicts of interest.


\bibliography{QW-Multigraphs-References}


\bibliographystyle{unsrt}

\end{document}